\documentclass[superscriptaddress,reprint,showpacs,amssymb,amsmath,aps,pra]{revtex4-1}
\usepackage{graphics}
\usepackage{bbold}
\usepackage{graphicx}
\usepackage{bm}
\usepackage{bbm}
\usepackage{amsmath,amsthm}
\usepackage{lipsum}
\usepackage{amssymb}
\usepackage{graphicx}
\usepackage{bm}
\usepackage{amsthm}
\usepackage{amsfonts}
\usepackage{color}
\bibpunct{[}{]}{,}{n}{}{}

\begin{document}
\title{Inducing critical phenomena in spin chains through sparse alternating fields}

\author{M. Cerezo}
\affiliation{Instituto de F\'{\i}sica de La Plata, CONICET, and Departamento de F\'{\i}sica,
Universidad Nacional de La Plata, C.C.\ 67, La Plata 1900, Argentina}
\author{R. Rossignoli}
\affiliation{Instituto de F\'{\i}sica de La Plata, CONICET, and Departamento de F\'{\i}sica,
Universidad Nacional de La Plata, C.C.\ 67, La Plata 1900, Argentina}
\affiliation{Comisi\'on de Investigaciones Cient\'{\i}ficas de la Provincia de Buenos Aires (CIC), 
La Plata 1900, Argentina}
\author{N. Canosa}
\affiliation{Instituto de F\'{\i}sica de La Plata, CONICET, and Departamento de F\'{\i}sica,
Universidad Nacional de La Plata, C.C.\ 67, La Plata 1900, Argentina}
\author{C. A. Lamas}
\affiliation{Instituto de F\'{\i}sica de La Plata, CONICET, and Departamento de F\'{\i}sica,
Universidad Nacional de La Plata, C.C.\ 67, La Plata 1900, Argentina}

\begin{abstract}
	We analyze the phase diagram  of the exact ground state (GS) of spin-$s$ chains with ferromagnetic $XXZ$ couplings under 
	{\it $n$-alternating} field configurations, i.e, sparse alternating fields having nodes at $n-1$ contiguous sites. 
	It is shown that  such systems  can exhibit a non-trivial magnetic behavior, which can differ significantly from that of the 
	standard ($n=1$) alternating case and  enable mechanisms for controlling their magnetic and entanglement properties. 
	The boundary in field space of the fully aligned phase can be determined analytically $\forall\,n$, and shows that it becomes reachable 
	only above a threshold value of the coupling anisotropy $J_z/J$, which depends on $n$ but is independent of the system size.  
	Below this value the maximum attainable magnetization becomes  much smaller. We then  show 	that the GS can exhibit significant 
	magnetization plateaus, persistent for large systems, at  which the  
	magnetization per site $m$ obeys the quantization rule $2n(s-m)=integer$,  consistent with 
	the Oshikawa, Yamanaka and Affleck (OYA) criterion. We also identify the emergence  of field induced  
	spin polymerization, which explains the presence of such plateaus. 
	Entanglement and field induced frustration effects are also analyzed.
	\end{abstract}

\maketitle
\vspace*{-1.2cm}

\section{Introduction}
\vspace*{-0.2cm}

One of the distinct hallmarks of cooperative behavior in interacting many-body quantum systems are the critical 
properties and phase transitions that arise when some control parameter is varied \cite{SS.99,LSM.61,MV.03,NL.05,ZW.18}.
In the last decades entanglement theory has  unveiled  new properties of these transitions, providing  a deep understanding  
\cite{OAFF.02,ON.02,VL.03, IJK.05, AFOV.08, ECP.10, SRP.14, NB.18}.  
In this scenario, the emergence of notable phenomena such as frustration \cite{LC.13,Sen.08,MM.10,Gil.11} and magnetization plateaus 
\cite{AH.04,AT.09,MT.11,CL.11,HH.14,OYA.97}, is typically associated with  antiferromagnetic systems with competing interactions 
\cite{MG.69,CL.16} and non-trivial geometries \cite{NS.13}. 
 However, much less is known of the critical properties that could be induced even in simple systems through general 
 non-uniform magnetic fields or couplings. Most investigations on nonuniform fields were focused so far on the alternating or  ``staggered''  case \cite{AL.95,GZ.05,MA.05,CRM.10,MA.11,ZV.18,ZS.16,CRC.15,TC.16}. Nonetheless,  recent studies with more general nonuniform fields  \cite{CRC.15,CRNR.17} 
 have shown that interesting and significant  phenomena can emerge, particularly with sparse field configurations \cite{CRNR.17}. 
 
 Interacting spin systems provide an adequate framework for studying  such  non trivial phenomena.  Moreover,  the possibility of 
 simulating spin  systems with tunable couplings and fields is  becoming increasingly feasible due to the recent remarkable advances in  
quantum control technologies \cite{LSA.12,IG.14,YS.16}.  
In particular, the paradigmatic $XXZ$ model 
\cite{SS.99,YY.66,JM.72,Al2,DK.02,SG.03,CR.06,JL.07,TL.13,JR.17,JR.18} can emerge as effective Hamiltonian in different systems \cite{BP.07,SO.11,XL.14,SS.15,AP.16,T.16,MV.16,SW.17,TN.18,MR.18,PC.04,BR.12,AP.16,BR.17}. For instance,  
it can be achieved in terms of superconducting charge qubits (SCQ) coupled with a SQUID (superconducting 
quantum interference device) \cite{XL.14,SS.15,MR.18}. In SCQ setups, the local field parameters can be 
controlled by means of a gate voltage applied to each SCQ box and an external magnetic flux is used to modulate 
the Josephson coupling energy \cite{XL.14}. Other examples comprise trapped ions \cite{IG.14,PC.04,BR.12,AP.16}, 
cold atoms in optical lattices  \cite{LSA.12,MV.16,SW.17,TN.18}, photon coupled microcavities \cite{BP.07},
 quantum dots \cite{SO.11}, etc. The $XXZ$ model  has also been employed for implementing quantum information protocols  \cite{LSA.12,IG.14,YS.16,SO.11,BB.04,BB.10}.

Here we will show that the application of sparse periodic alternating fields in a ferromagnetic $XXZ$ system of arbitrary spin results 
in novel ground state (GS) phase diagrams, which display non-trivial magnetization plateaus and entanglement properties.  In the first place, 
the boundary in field space of the fully aligned phase, which determines the onset of GS entanglement, 
can be determined analytically and implies a threshold value of the coupling anisotropy, 
below which the maximum attainable magnetization becomes much smaller. Such boundary is independent of the system size. 
 It is then shown that such sparse fields can induce other non-trivial magnetization plateaus, persistent for large sizes, 
 as verified through DMRG \cite{SW.93,AK.96,US.03} calculations. 
These plateaus are shown to satisfy the well known OYA criterion \cite{OYA.97}, 
which can be here explained simply through field induced polymers with definite magnetization. 
 We also analyze other aspects like field induced frustration, single-spin magnetization and pairwise 
 entanglement, whose results support the  polymerization based picture.  
 
The model and the $n$-alternating field configuration are described in sec.\  \ref{II}, with the 
boundary of the fully aligned phase and the conditions under which it can be reached discussed in 
\ref{IIA}. GS magnetization diagrams are then discussed in \ref{IIB}, while  pairwise entanglement  in  \ref{IIC}.
  The appendices contain the derivation of analytic expressions for the previous boundary and for entanglement
   measures at the boundary, and the exact analytic solution  of the limit case of an XX chain 
  under the present field configurations.  Conclusion are  drawn in \ref{III}. 

\section{Sparse alternating field configurations\label{II}}
We consider a cyclic chain of $N$ spins $s$ interacting through first-neighbor $XXZ$ couplings in  
a non-uniform magnetic field along the $z$ axis. The Hamiltonian reads
\begin{equation}
H=-\sum_{i=1}^N[ h^i S_i^z+J(S^x_i S^x_{i+1}+S^y_i S^y_{i+1})+J_zS^z_i S^z_{i+1}]\,, \label{H1}
\end{equation}
where $h^i$, $S_i^\mu$ are the field and spin components at site $i$ (with $N+1\equiv 1$) and $J$, $J_z$ the coupling strengths. 
As $[H,S^z]=0$, with $S^z=\sum_j S^z_j $ the total spin along the $z$ axis, its eigenstates can be characterized 
by the total magnetization $M=S_z$ ($-Ns\leq M\leq Ns$). We will set $J>0$, as the spectrum and entanglement properties of $H$ 
are the same for $\pm J$ \cite{expl1}. They are also identical for $(\{h^j\},M)$ and $(\{-h^j\},-M)$ \cite{expl2}. 
It is as well convenient to use the scaled coupling strengths 
\begin{equation}
    j_z=2sJ_z,\;\;j=2sJ\,,\label{jzj}
\end{equation}
 as critical fields and couplings will depend just on $j_z$ and $j$ for different values of $s$  (see below). 
 
\begin{figure}[t]
  \centering{{\includegraphics[width=.8\columnwidth]{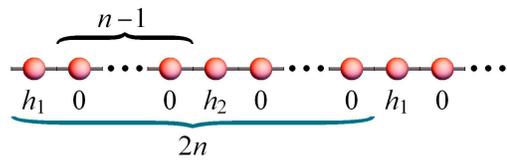}}} 
\caption{Schematic representation of a spin chain with an $n$-alternating field configuration (Eq.\ (\ref{naf})). 
 The number of intermediate sites with no field is $n-1$. The period is $2n$.}
  \label{f1}
\end{figure}

 We will here examine  the {\it $n$-alternating field configuration}, depicted in Fig.\ \ref{f1}, defined by
\begin{equation}
  h^i=\left\{
  \begin{array}{@{}ll@{}}
    h_1, & i=1,2n+1,4n+1,\ldots\\
    h_2, &i=n+1,3n+1,5n+1,\ldots\\
     \,0, & \text{otherwise} 
  \end{array}\right.\;,\label{naf}
\end{equation}
which generalizes the standard alternating (A)  case  $(h_1,h_2,h_1,h_2,\ldots)$,  recovered for $n=1$. For $n=2$ we obtain 
the ``next-alternating'' (NA) case    $(h_1,0,h_2,0,\ldots)$, while for $n=3$  the  
``next-next-alternating'' (NNA) case $(h_1,0,0,h_2,0,0,\ldots)$. We set in what follows $N=2nK$, with $K$  the number of cells with $2n$ spins. 

A  motivation for studying the field configurations (\ref{naf}) in the present system is that for $j_z>j>0$, they {\it all}  
exhibit, for {\it any} spin $s\geq 1/2$, a  {\it multicritical} point in the GS at fields of opposite sign given by \cite{CRNR.17}
\begin{equation} 
h_1=-h_2=\pm h_s,\;\;h_s=\sqrt{j_z^2-j^2}\,,\label{hs}
\end{equation}
where {\it all GS magnetizations plateaus merge}: At this point the GS becomes $2Ns+1$ degenerate, with the  GS's 
for each magnetization $M$ having all the same energy.  This point generalizes the Pokrovsky-Talapov
(PT)-type transition  of a spin-1/2 chain in an alternating
field \cite{AL.95}. Furthermore, at this point there is a whole family of completely separable {\it factorized} 
(i.e. product) exact GS's \cite{CRNR.17}, and  the field (\ref{hs}) is then denoted as  {\it factorizing} (or separability)  
field \cite{CRNR.17}. It is independent of the chain size $K$ and the distance $n$ between spins with field, depending 
just on the scaled couplings (\ref{jzj}). Here we will show that the field configurations (\ref{naf}) exhibit other 
interesting properties in the present system for $j_z<j$, being  capable of inducing a non-trivial magnetic response. 

\subsection{Border of the aligned phase \label{IIA}}

A first basic question which arises for $n\geq 2$ is if such sparse fields are sufficient to induce a completely 
 aligned (and hence completely separable) GS with maximum magnetization $|M|=Ns$. In the standard staggered case $n=1$, such phase will always arise 
for sufficiently strong fields $h_1,h_2$ of the same sign, for any value of $j$ or $j_z$,  but for $n\geq 2$ 
 the presence of spins with zero field implies  that it will not be attainable without the aid of a finite positive value of $j_z$, as shown below. 

We will prove in fact that for $n\geq 2$, the aligned phase with $|M|=Ns$ is attainable only for 
\begin{equation}
   j_z>j_{z}^c(n)=j\cos(\pi/n)\,,\;\;n\geq 2\,,\label{critz}
\end{equation}
in which case the GS will have $M=Ns$ if the fields $(h_1,h_2)$ satisfy $h_1+h_2>0$ and 
\begin{equation}
(h_1+\beta_n)(h_2+\beta_n)>\alpha_n^2\,,\label{crit}
\end{equation} 
with $h_i>-\beta_n$ such that $(h_1,h_2)$  lies above the upper branch of the hyperbola, and $M=-Ns$ if $h_1+h_2<0$ and
\begin{equation}
(h_1-\beta_n)(h_2-\beta_n)>\alpha_n^2\,,
\label{crit2}
\end{equation}
with $h_i<\beta_n$ such that $(h_1,h_2)$  lies below the lower branch of a reflected hyperbola. The coefficients $\alpha_n$, $\beta_n$ are  
independent of the size $K$ and are given by  
\begin{eqnarray}
\alpha_n=j\frac{\sinh \gamma}{\sinh n\gamma}&=&2h_s\frac{j^{n}}{(j_z+h_s)^n-(j_z-h_s)^n}\,,\label{alph}\\
\beta_n=j\frac{\sinh\gamma}{\tanh n\gamma}&=&
h_s\frac{(j_z+h_s)^n+(j_z-h_s)^n}{(j_z+h_s)^n-(j_z-h_s)^n}\,,
\label{beta}
\end{eqnarray}
where $\cosh\gamma=j_z/j=\Delta$ is the anisotropy and  $h_s=j\sinh \gamma$  the separability field (\ref{hs}), satisfying $\beta_n^2=\alpha_n^2+h_s^2$. 
Eqs.\ (\ref{alph})-(\ref{beta})  hold and are real for both  $j_z>j$, where $\gamma$ and $h_s$ are real, and also   
$\;j_z^c(n)<j_z<j$, where $\gamma$ and  $h_s$ become imaginary: $\gamma=\imath\phi$, with $j_z/j=\cos\phi$, $h_s=\imath j\sin\phi$ and 
\begin{equation}\alpha_n=j\frac{\sin\phi}{\sin n\phi}\,,\;\;\;\;\beta_n=j\frac{\sin\phi}{\tan n\phi}\,,\;\;\;\;(j_z<j)\label{al2}\,.
\end{equation}
In the isotropic limit $j_z\rightarrow j$, $\phi\rightarrow 0$ and $\alpha_n=\beta_n\rightarrow j/n$. 

{\it Proof.} The boundary in field space $(h_1,h_2)$ of the fully aligned phase can be obtained by  determining the fields at which the GS undergoes the magnetization 
transition $|M|=Ns\rightarrow Ns-1$, i.e., 
where the fully aligned state starts to become unstable against single spin excitations.  
The fully aligned states  $|M=\pm Ns\rangle$ are trivial eigenstates of $H$ $\forall\,n$ in (\ref{naf}), with energies  
\begin{equation}E_{\pm Ns}
=-Ks[\pm(h_1+h_2)+nj_z],\label{Eal}\end{equation}  which are independent of $j$ and degenerate for $h_1+h_2=0$. They will be the GS for sufficiently large $j_z$ 
and/or  strong positive ($M=Ns$) or negative ($M=-Ns$) fields. 

On the other hand,  the $|M=Ns-1\rangle$ eigenstate of lowest energy can be obtained by diagonalizing $H$ 
in the invariant subspace spanned by the $2n$ $W$-like \cite{WD.00}  states  with one spin down (here $S^-_i=S^x_i-iS^y_i$), 
\begin{equation}|W_i\rangle=\frac{1}{\sqrt{2sK}}\sum_{l=0}^{K-1}
S_{i+2nl}^{-}|Ns\rangle,\;\;\;\;i=1,\ldots,2n\,,\label{W}
\end{equation} 
where all sites with the same position $i$ in the  cell have the same weight.  These states lead  to  close and size-independent matrix elements 
of $\Delta H=H-E_{Ns}$: 
\begin{equation}
\Delta H|W_i\rangle=(j_z+h^i)|W_i\rangle-\eta_nj(|W_{i+1}\rangle+|W_{i-1}\rangle)
\label{H_n1}
\end{equation}
where $h^i=\delta_{i1}h_1+\delta_{i,n+1}h_2$, $\eta_n=1$ ($1/2$) for $n=1$ ($\geq 2$) and $|W_0\rangle=|W_{2n}\rangle$, $|W_{2n+1}\rangle=|W_{1}\rangle$. 
A stable $M=Ns$ GS requires $\Delta H$ {positive definite}, entailing positive eigenvalues  (excitation energies) of the $2n\times 2n$ matrix $\Delta H_n$ of elements 
\begin{equation}
(\Delta H_n)_{ij}=\langle W_i|\Delta H|W_j\rangle=\delta_{ij}(j_z+h^i)-\eta_n j\delta_{i,j\pm 1}\,.\label{Hn}
\end{equation}
 This implies the necessary condition 
\begin{equation}
{\rm Det}\,[\Delta H_n]>0\,. \label{Det}
\end{equation}
Assuming $\Delta H$ positive definite for strong positive fields,  the $M=Ns\rightarrow Ns-1$ transition then occurs at fields $(h_1,h_2)$ 
which are the first root of ${\rm Det}[\Delta H_n]=0$ when approached from the strong positive field limit. From Eq.\ (\ref{Hn}) it is seen that this determinant has the form 
\begin{eqnarray} {\rm Det}\,[\Delta H_n]&=&a_nh_1h_2 +b_n(h_1+h_2)+c_n\label{dan}\\
&=&a_n[(h_1+\beta_n)(h_2+\beta_n)-\alpha_n^2]\,,\label{dan2}
\end{eqnarray}
with $\beta_n=b_n/a_n$, $\alpha_n^2=\beta_n^2-c_n/a_n$ and $a_n,b_n,c_n$ field-independent. 
Their  expressions (\ref{alph})--(\ref{beta}) are derived in Appendix A, where it is shown that $a_n\geq 0$ (Eq.\ (\ref{an})). 
Then, positivity of $\Delta H$ implies  fields $(h_1,h_2)$ satisfying (\ref{crit}), with $h_i>-\beta_n$. And stability with respect to the $M=-Ns$ GS requires $h_1+h_2>0$ 
(Eq.\ (\ref{Eal})). 
 A similar procedure shows that an aligned GS with $M=-Ns$ requires  fields satisfying  (\ref{crit2}) with $h_i<\beta_n$ and $h_1+h_2<0$. 

As $j_z/j$ decreases below $1$, the denominators in (\ref{al2}) become smaller,  {\it vanishing} for $\phi\rightarrow \pi/n$ if $n\geq 2$, i.e., 
for $j_z$ approaching the {critical value} (\ref{critz}). This implies the {\it divergence} of $\alpha_n$ and $\beta_n$, and hence of the critical 
fields, in this limit (note that $\beta_n<0$  for $j_z^c(n)<j_z<j_z^{c}(2n)$).  
The fully aligned phase becomes then {\it unreachable for $j_z\leq j_{z}^c(n)$} ($\phi\geq \pi/n$). 
This result can also be directly derived from (\ref{Hn}): As shown in Appendix A, the lowest eigenvalue $\lambda_0(n)$ of the matrix $\Delta H_n$ satisfies
\begin{equation} \lambda_0(n)< j_z-j\cos(\pi/n)\,,\end{equation}
with the upper bound reached for $h_1,h_2\rightarrow+\infty$.  
Thus, for $j_z\leq j\cos(\pi/n)$, $\Delta H_n$ has a negative eigenvalue at  {\it all} finite fields and the aligned state cannot be a GS. 
It is also verified that  at the critical value  $j_z=j\cos(\pi/n)$, $a_n$, $b_n$ and $c_n$  in Eq.\ (\ref{dan}) {\it vanish},  i.e., ${\rm Det}\,[\Delta H_n]=0$ 
$\forall$ ($h_1,h_2)$ (see Appendix A).\qed 

The hyperbolas which delimit the aligned phase  in Eqs.\ (\ref{crit})--(\ref{crit2}) also represent  {\it the onset of GS entanglement}, 
and correspond to an {\it entanglement transition}:  The $|M=Ns-1\rangle$ GS will be 
of the form $\sum_{i=1}^{2n} w_i |W_i\rangle$, with $w_i>0$ and $\sum_{i=1}^{2n}w_i^2=1$, which is an entangled state.

 Due to the form (\ref{W}) of the states $|W_i\rangle$,  
pairwise entanglement will reach {\it full range} in this sector, since 
the ensuing reduced state of two spins, $\rho_{ij}$, will depend just on their positions $i,j$  within the cell 
{\it but not on their distance}, i.e., on the number of cells between them. Since $\rho_{ij}$ 
is a mixed state, its entanglement can be measured through the entanglement of formation $E_f(\rho_{ij})$ \cite{BD.96}, 
defined as the convex roof extension of the pure state entanglement entropy.  Moreover, as the present $\rho_{ij}$ can be 
considered as an effective  two-qubit state, $E_f(\rho_{ij})$  can be determined analytically by means of the concurrence 
\cite{WW.97} $C_{ij}=C(\rho_{ij})$, which is itself an entanglement measure \cite{GV.00}, with $C_{ij}=1$ $(0)$ for a maximally entangled (separable) mixed state 
(see Appendix B for details and precise definitions of these quantities).   
 As $\rho_{ij}$ is independent of the distance between the spins, so is the pairwise concurrence, which is given by  (see again Appendix B) 
\begin{equation} C_{ij}=2|w_iw_j|/K\,.\label{Cij}\end{equation} 
This value  {\it saturates} the monogamy  relations \cite{CKW.00,OV.06}. 

On the other hand, at the mean field  level  the separable fully aligned states are the trivial symmetry preserving mean field solutions, 
and the hyperbolas in Eqs.\ (\ref{crit})--(\ref{crit2}) represent the onset of the {\it symmetry-breaking} mean field phase, i.e.\ of degenerate mean field 
solutions with $\langle S^\mu_i\rangle\neq 0$ for $\mu=x$ or $y$ (see also Appendix A). For $j_z\leq j_z^c(n)$ the aligned solutions are unstable at all fields.  
 
 We finally  remark that for $j_z<j_z^c(n)$, 
 the instability of the aligned state also holds at the single cell level, entailing that a whole interval of  
 magnetizations (at least $Ns-K+1\leq M\leq Ns$) also cease to be stable,  as will be verified in the next section. 

\begin{figure}[t]
 \hspace*{-0.3cm}{\includegraphics[width=1.05\columnwidth]{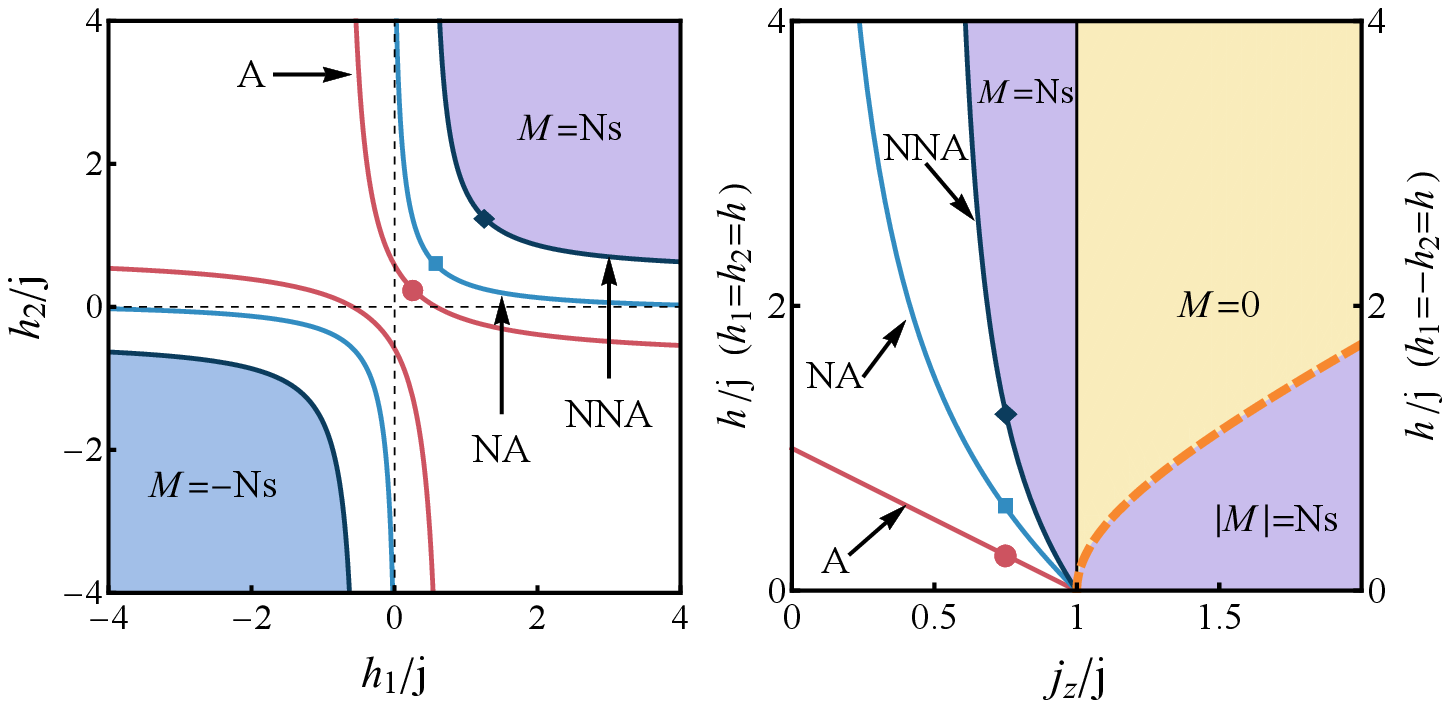}}
\caption{GS phase diagrams for spin-$s$ $XXZ$ chains in the alternating ($n=1$, A),  
	``next-alternating'' ($n=2$, NA) and ``next-next alternating'' ($n=3$, NNA)  field configurations. 
	Left: The hyperbola branches of Eqs.\  (\ref{crit})--(\ref{crit2}) delimiting the fully aligned $M=\pm Ns$ phases at  $j_z/j=0.75$. 
 Colored regions indicate GS magnetization $|M|=Ns$ for all three  cases. Right: The threshold  (\ref{hcn})  (solid lines)  
  of the GS aligned phase for parallel fields $h_1=h_2=h$ and  $j_z<j$,  which diverges for $j_z\rightarrow 0$ ($j/2)$ 
 in the NA (NNA) case, and its upper limit (\ref{hs}) (dashed line) for antiparallel fields $h_1=-h_2=h$ and $j_z>j$, 
 common for all $n$.  Points indicate the thresholds for $h_1=h_2$ at $j_z/j=0.75$.}\label{f2}
\end{figure}

{\it The first three cases}. Let us now examine the particular cases $n=1$, $2$ and $3$ in Eq.\ (\ref{naf}).   
In the standard staggered case $n=1$, Eqs.\ (\ref{alph})--(\ref{al2}) lead to  
\begin{equation}\alpha_1=j,\;\;\;\beta_1=j_z\,,\label{cz}\end{equation}  
being then verified from (\ref{crit})--(\ref{crit2})  that the aligned phase is
 reachable $\forall$ $j,j_z$ for sufficiently strong $h_1,h_2$. 
However, in the NA case $n=2$, they imply   
\begin{equation}
\alpha_2=\frac{j^2}{2j_z}\,,\;\;\;\beta_2=j_z-\frac{j^2}{2j_z},\;\;\;\;j_z>0,
\label{crit22}     \end{equation}
which {\it diverge} for $j_z\rightarrow j_z^c(2)=0$. Increasingly stronger fields are here required to reach the aligned phase as $j_z$ decreases, 
diverging in the $XX$ limit $j_z=0$. 
For  $j_z\leq 0$ it becomes  unreachable (see also Appendix C). 

\begin{figure}[t]
  	\hspace*{-0.25cm}{\includegraphics[width=1.05\columnwidth]{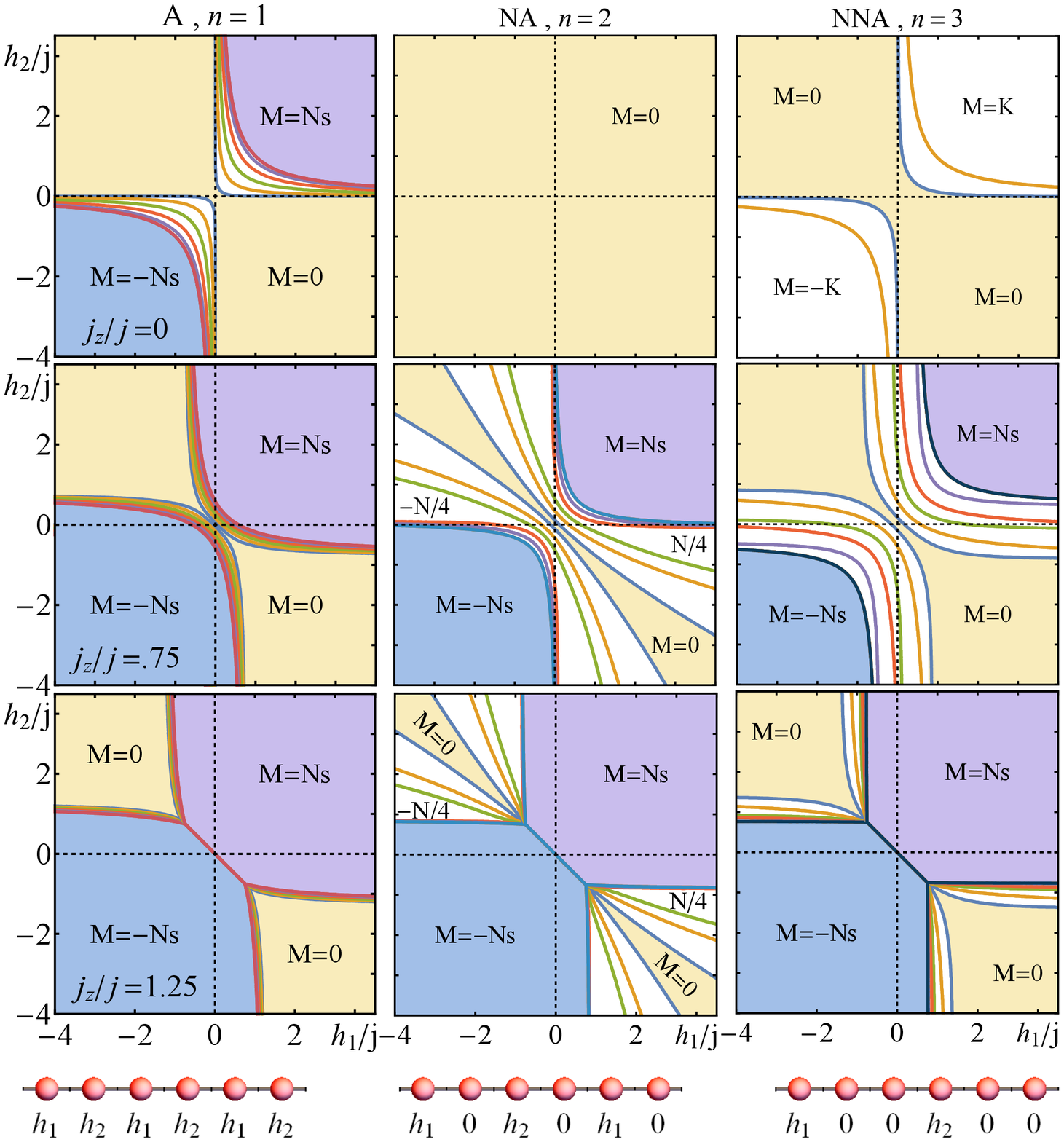}}
  	\caption{GS magnetization diagram in the ($h_1,h_2$) field plane for an $N=12$ spin-$1/2$ $XXZ$ chain 
  		in an $n=1$ (left), $n=2$ (center) and $n=3$ (right) field configuration  (\ref{naf}), for anisotropies $j_z/j=0$ (top), $0.75$ 
  		(center) and $1.25$ (bottom).  Curves separate different  magnetizations. 
  		For $j_z=0$ the GS reaches all magnetizations $-N/2\leq M\leq N/2$ in the A  case,  but remains at $M=0$ $\forall$ $(h_1,h_2)$ in NA, 
  		and reaches just the $M=\pm K=\pm 2$ plateaus in NNA fields. For $j_z/j=0.75$, the three configurations reach all magnetizations, 
  		as now $j_z>j_z^c(n)$, although the NA case stands  out for its wide $|M|\leq N/4$ sectors. For $j_z>j$, magnetizations $|M|<N/2$ 
  		are reached only for fields of opposite sign beyond the factorization points $h_1=-h_2=\pm h_s$, independent of $n$. Bottom row: 
  		schematic representation of field configurations.}  \label{f3}
  \end{figure}

And in the NNA case $n=3$, Eqs.\ (\ref{alph})--(\ref{al2}) lead to 
\begin{equation}
\alpha_3=\frac{j^3}{4j_z^2-j^2},\;\;
\beta_3=j_z\frac{4j_z^2-3j^2}{4j_z^2-j^2},\;\;\;\;\;j_z>j/2,
\label{crit33}    
\end{equation}
which diverge already for $j_z\rightarrow j_z^c(3)=j/2$.  The aligned GS cannot be reached  for  $j_z\leq j/2$.
 The critical fields and couplings of these  three cases are depicted in Fig.\ \ref{f2}, with the GS magnetization diagrams shown in Fig.\ \ref{f3}. 

{\it The  parallel critical field.}  Eqs.\ (\ref{crit})--(\ref{al2}) also entail  that  if  $j_{z}^c(n)<j_z<j$, 
full alignment requires application of non-zero fields. For $h_1=h_2=h$, they imply 
\begin{equation}|h|>h_c^{\parallel}(n)=\alpha_n-\beta_n=j\sin\phi\tan\tfrac{n\phi}{2}
\,,
\label{hcn}\end{equation}
which defines a parallel critical field $h_c^{\parallel}(n)$. And if $h_2=0$, a single field $|h_1|>-h_s^2/\beta_n=h_c^{||}(2n)$ is sufficient 
provided $\beta_n>0$, i.e.\ $\phi<\frac{\pi}{2n}$, which is equivalent to $j_z>j_z^c(2n)$ ($h_2=0$  in the  $n$-alternating configuration is 
equivalent  to  $h_2=h_1$ in the  $2n$-alternating case). 

In contrast, for $j_z>j$ the GS is fully aligned already at zero field $\forall\, n$ and lower magnetizations $|M|$ arise 
only for fields of {\it opposite} sign beyond the factorizing  
points $h_1=-h_2=\pm h_s$ \cite{CRNR.17},  where {\it all} magnetization plateaus coalesce, 
as seen in the bottom panels of Fig.\ \ref{f3}.  The upper and lower branches of the hyperbolas 
 (\ref{crit})--(\ref{crit2}) intersect precisely at these points $\forall$ $n$,  providing the border 
 of the aligned phase just beyond these points.  Between them, the aligned phases touch 
 at the line $h_1+h_2=0$.  Note also that for $j_z>j$ and $n>1$, $\alpha_n$ becomes rapidly 
 small for large $n$ ($\alpha_n\approx 2h_s(\frac{j}{j_z+h_s})^n$) or large $\frac{j_z}{j}$ 
 ($\alpha_n\approx j(\frac{j}{2j_z})^{n-1}$), implying $\beta_n\approx h_s$ and  hence 
non alignment ($|M|<Ns$) just for 
$|h_i|\agt h_s$ for $i=1,2$ (and $h_1h_2<0$). 

When formally extended to all values of  $j_z$, the antiparallel (\ref{hs}) and parallel  (\ref{hcn}) critical fields  
fully determine $\alpha_n$ and $\beta_n$, and hence the whole border of the aligned phase: 
 \begin{equation} \begin{array}{c}\alpha_n\\\beta_n\end{array}=\tfrac{1}{2}[\pm 
 h_c^{||}(n)-h_s^2/h_c^{||}(n)]\,,\end{equation} 
 where $h_c^{\parallel}(n)=-j\sinh\gamma\tanh \frac{n\gamma}{2}$ for $j_z>j$. For $j_z\rightarrow j$ 
 both vanish but $-h_s^2/2h_c^{\parallel}(n)\rightarrow j/n$.
  
 \subsection{Magnetization \label{IIB}}
\begin{figure}[t]
 \hspace*{-0.25cm}{\includegraphics[width=1.05\columnwidth]{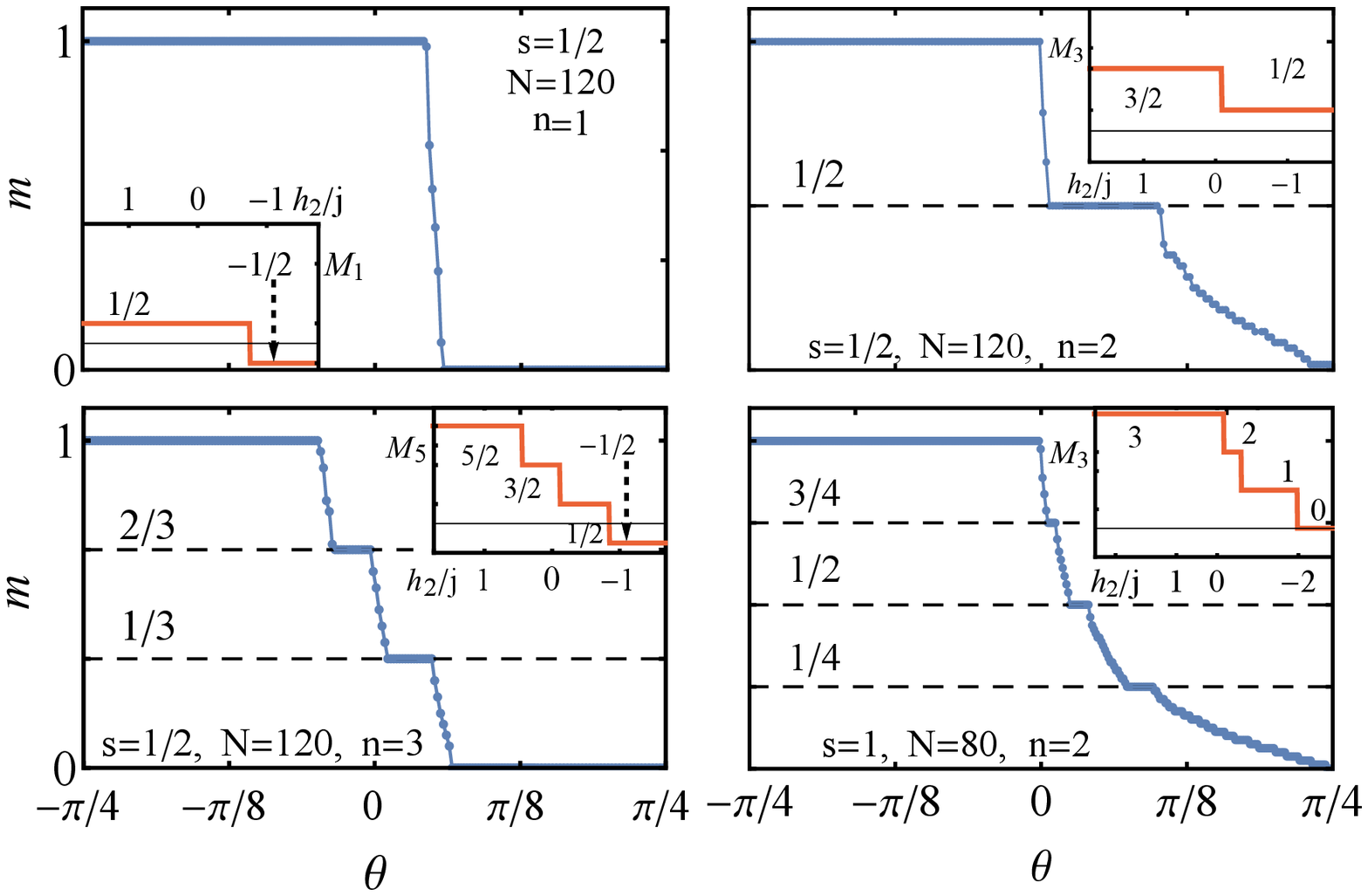}}
\caption{Scaled GS magnetization $m=M/(Ns)$ for 
$j_z/j=0.75$ and fields $(h_1,h_2)=4j(\cos\theta,-\sin\theta)$. Results for 
$s=1/2$,  $N=120$ spins and $n=1,2,3$, and for $s=1$, $N=80$ spins and $n=2$, are depicted.  
 Insets show the polymer magnetizations $M_{2n-1}$.}  \label{f4}
\end{figure}

 A second fundamental question which arises is if magnetization plateaus with $|M|<Ns$ of {\it significant width} do also emerge.  
 For large systems the GS will indeed possess such  plateaus (Fig.\ \ref{f4}), at which the scaled magnetization 
 $m=M/(Ns)$ obeys the quantization rule
\begin{equation}
2ns(1-m) = q\,,
\label{OYA}
\end{equation}
with $q$ integer. This result can be readily understood by considering the situation where one of the fields ($h_1$) is sufficiently 
strong so that the spin chain can be viewed approximately as $K$ {\it polymerized subsystems} consisting of  $2n-1$ spins-$s$ with a 
field $h_2$ at the central site (Fig.\ \ref{f1}), separated by fully aligned spins. When $h_2$ is varied the polymer GS 
magnetizations $M_{2n-1}$ will be $(2n-1)s-q$ with $q$ integer, starting from $q=0$ when $j_z>j_z^c(n)$. Therefore, the total 
GS magnetization will be $K\left\{s+[(2n-1)s-q]\right\}$, entailing then (\ref{OYA})  and meaning that {\it the plateaus in $m$ 
reflect essentially the polymer magnetizations}. Due to the periodicity,  Eq.\ (\ref{OYA}) is consistent with the OYA criterion \cite{OYA.97}   
(normally used in antiferromagnetic chains in uniform fields). 
Intermediate magnetizations arise  then in the transition regions between these plateaus 
and imply no definite magnetization at the single cell level.

In Fig.\ \ref{f3} we show representative results for the GS magnetization in a small spin $1/2$ chain. 
In the standard alternating case $n=1$ (left panels), the GS reaches {\it all} magnetizations for any anisotropy $j_z/j$, 
with the fully  aligned $|M|=N/2$ sectors separated from the $M=0$ plateau by a narrow band containing all intermediate magnetizations. 
In contrast, in an $n=2$ NA configuration (center), it is first verified that for $j_z=j_z^c(2)=0$, the GS cannot be fully aligned. 
Moreover, {\it it has strictly $M=0$ for all fields}, as can be rigorously shown through its Jordan-Wigner 
fermionization \cite{LSM.61} (see Appendix C). 
And for $j_z>0$ this configuration exhibits a noticeable behavior, showing  wide $0\leq |M|\leq N/4$ sectors in addition to  
the aligned phases, with the $|M|=N/4$ plateau persisting for large $N$ (see below). Finally, in the NNA  $n=3$ case (right),  
it is again verified that if $j_z\leq j_{z}^c(3)=j/2$, 
the GS cannot be fully aligned (top panel),  reaching instead a maximum  magnetization  $|M|= 2sK=N/6$ for $j_z=0$ 
(and also $0<j_z<j/2$ if $s=1/2$): For strong parallel fields, spins with field become aligned while  those without   
form essentially {\it entangled  
dimers} with zero magnetization, entailing  $|M|=2sK$.  And when $j_z>j/2$, the magnetization 
diagram becomes similar to that of the $n=1$ case, 
although with a much wider transition sector between the $M=0$ and $|M|=N/2$ plateaus. 

Previous results imply  that the threshold $j^c_z(n)$ of the aligned phase is actually a {\it critical point} 
below which a whole interval of 
magnetizations cease to be reachable. This can be understood again from the strong field limit 
$h_1,h_2\rightarrow \infty$, 
where spins with field are fully aligned while those without form essentially isolated chains of $n-1$ spins, 
with effective fields $sJ_z$ at the endpoints: 
For $n=2$ and $j_z\rightarrow 0^+$,  all magnetizations $M\geq 0$ (and not just $N/2$ and $N/2-1$) of the whole  
chain become degenerate at strong fields, 
since the $M_1=\pm 1/2$ states of each of the $2K$ single spins without field become degenerate, 
remaining just $M=0$  for $j_z\leq 0$.  
Similarly, for $n=3$ and $j_z\rightarrow j/2$, all chain magnetizations $M\geq 2sK=N/6$ become degenerate  at strong fields, 
since each pair without field may have magnetizations $M_2=1$ or $0$, degenerate precisely at $j_z=j/2$.

In Fig.\ \ref{f4} we show the GS scaled magnetization for a chain of $N=120$ spins-$1/2$, obtained with density  matrix 
 renormalization (DMRG) \cite{SW.93,AK.96,US.03,DMRG}. In the $n=1$ case the transition region from $M=0\rightarrow  N/2$ is 
 again quite narrow (top left), in agreement with (\ref{OYA}), since here the ``polymer'' formed for large $h_1$
consists of just one spin-$s$, whose lower state
may have only two magnetizations: $M_1=s$ and $-s$ (see inset), i.e.\ $q=0$ and $q=2s$, leading just to $|m|=1,0$ plateaus. 
For $n=2$ and $j_z>0$, the GS possesses plateaus at $|m|=1,1/2$ (top right), reflecting the magnetizations $M_3=3/2,1/2$ ($q=0,1$) 
of the trimer formed by the three spins trapped between two aligned spins. 
Moreover, the trimer cannot reach $M_3=-1/2$ (except for large  $h_2\approx -h_1$) entailing no wide  $m=0$ plateau. For $n=3$, 
however, pentamer magnetization $M_5$ does reach $-1/2$, entailing a large $m=0$ plateau, in addition to the aligned phase $|m|=1$ 
($M_5=5/2$, $q=0$) and smaller intermediate plateaus at $|m|=2/3,1/3$ ($M_5=3/2,1/2$, $q=1,2$, bottom left). Such persistent plateaus 
also occur for higher spins, as seen for $s=1$ and $n=2$ (bottom right), 
where $|m|=1,3/4,1/2,1/4$, following the trimer magnetizations $M_3=3,2,1,0$. 

\begin{figure}[t]
 \hspace*{-0.5cm}{\includegraphics[width=1.1\columnwidth]{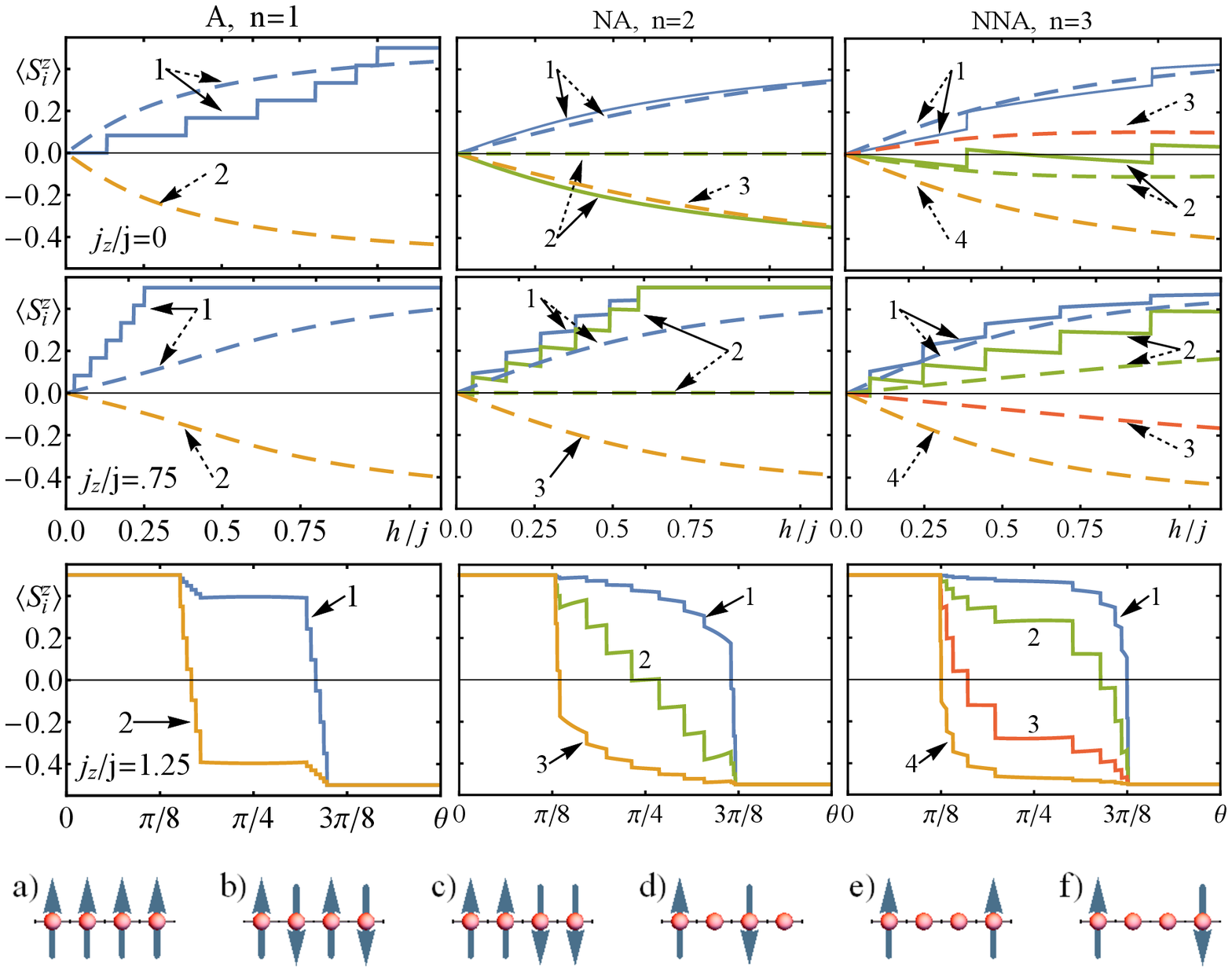}}
\caption{Top: Individual magnetizations $\langle S^z_i\rangle$ of the first four spins in the same chains of  
	Fig.\ \ref{f3}. 
	In the first and second row solid lines depict magnetizations for parallel 
	fields $h_1=h_2=h$, and dashed lines for 
antiparallel fields $h_1=-h_2=h$, while in the third row fields are selected as    
$(h_1,h_2)=2j(\cos\theta,-\sin\theta)$, with $\theta \in (0,\pi/2)$. 
The steps in $\langle S^z_i\rangle$ reflect those of the total magnetization $M$.  Bottom: schematic 
representation of spin configurations. }  \label{f5}
\end{figure}

Fig.\ \ref{f5} shows the single spin magnetization $\langle S^z_i\rangle$ of the first four spins in the chains  of Fig.\ \ref{f3}. 
For $s=1/2$, $1/2-|\langle S^z_i\rangle|$ is also a measure of the {\it entanglement} of spin $i$ with the rest of the chain 
(i.e., of the mixedness of the single spin reduced state \cite{expl3}), with $|\langle S^z_i\rangle|=0$ ($1/2$) 
implying maximum (zero) $i-$rest entanglement. 

The spins with field will align with the field direction as $h_i$ increases, 
leading for $n=1$ to type-a (b) spin configurations for strong parallel (antiparallel) fields. However, 
those without field ($n\geq 2$) exhibit a more complex behavior. 
For $n=2$ and $j_z=0$, the total GS magnetization $M$ vanishes $\forall\,h_1,h_2$,  implying that these spins 
become {\it antialigned} for $h_1=h_2$, leading to a type-b N\'eel configuration, but have {\it zero magnetization} 
($\langle S^z_i\rangle=0$) for $h_1=-h_2$, entailing a type-d configuration. This configuration also holds for $j_z>0$ if $h_1=-h_2$ 
(and $|h_i|>h_s$ if $j_z>j$), since $M$ still vanishes, implying that these spins become {\it frustrated}, 
as the attractive $S^z_i S^z_{i+1}$ coupling cannot be satisfied with both adjacent spins. This is a clear example 
of {\it field-induced frustration}, and entails {\it maximum $i$-rest entanglement}, mostly saturated with 
neighboring zero field spins. On the other hand, for large $h_1=h_2$ and $j_z>0$, they become aligned (type-a). 
 
In contrast, for $n=3$ the two contiguous spins without field tend to form an {\it entangled dimer}, leading for $j_z=0$ 
to a type-e configuration ($\langle S^z_i\rangle\approx 0$ for $i=2,3$) if $h_1=h_2$ and a type-f 
configuration if $h_1=-h_2$, here slightly polarized towards b. In this case there is actually a 
spin configuration transition when $0<j_z<j_z^c(3)=j/2$,  where  $\langle S^z_i\rangle$ changes sign at the central spins  
and the polarization evolves from type-b to type-c, crossing exactly type-f. 
For $j_z>j/2$, these central spins remain significantly entangled for antiparallel fields, 
 polarized towards type-c, while for parallel fields they become increasingly aligned as $|h_i|$ and hence $|M|$ increases. 
 Previous behaviors can also be seen at the bottom panels 
 for $j_z>j$,  which depict the ``evolution'' of $\langle S^z_i\rangle$ with $\theta=\tan^{-1}(-h_2/h_1)$ 
 between the fully aligned phases. 

\subsection{Pairwise Entanglement\label{IIC}}

\begin{figure}[ht!]
	\centering{{\includegraphics[width=1.05\columnwidth]{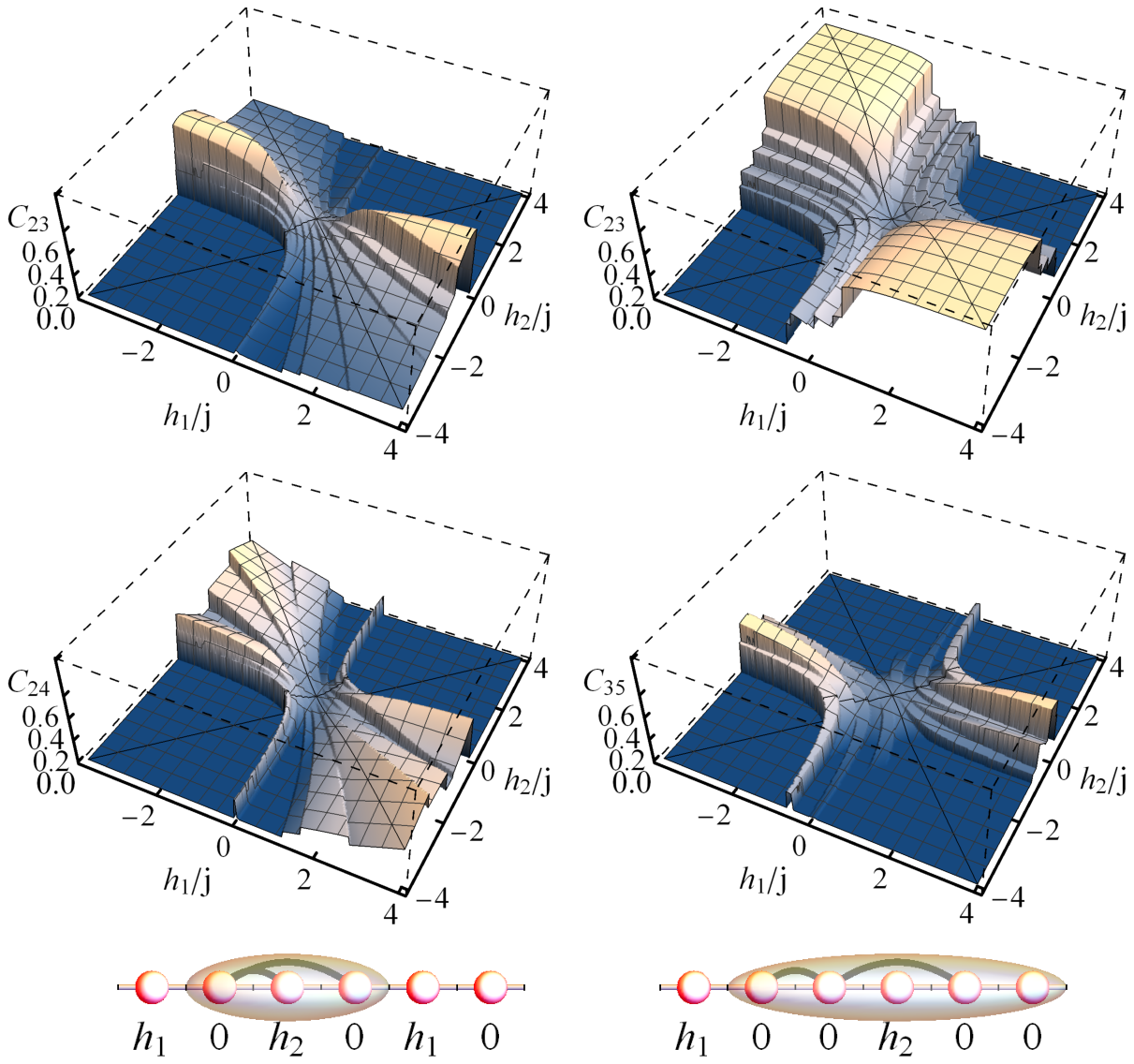}}} 
	\caption{Concurrence $C_{ij}$ between spins $i$ and $j$ (joined by a line in the bottom row) in the $(h_1,h_2)$ field space 
		for the exact GS of an $N=12$, $s=1/2$ chain with $j_z/j=0.75$,   in   NA (left) and NNA (right) field configurations.  
		The steps reflect the different total chain magnetizations. The onset of entanglement is determined 
		by the border of the aligned phase.
		Bottom: Schematic representation of approximate trimerization occurring in the $\pm N/4$ plateaus for $n=2$, and 
		pentamerization in the $\pm N/3$ and $\pm N/6$ plateaus for $n=3$. }  \label{fs}
\end{figure}

We show in Fig.\ \ref{fs} illustrative results for the pairwise entanglement  measured through the concurrence 
\cite{WW.97}, in the chains of Fig.\ 3  for $j_z/j=0.75$.  
It is first verified that in the $n=3$ NNA case,  the   two contiguous spins with zero field ($C_{23}$, top right) 
are highly entangled in the $M=0$ plateau, since the  spins form there essentially a type-f dimerized configuration 
(see bottom row of Fig.\ 5). Accordingly, the concurrence $C_{35}$ of a non-contiguous pair with zero field spins 
(bottom right) vanishes in this plateau. In contrast, the latter becomes significant in the $|M|=4$ and $|M|=2$ 
plateaus ($|m|=2/3, 1/3$), where the intermediate field $h_2$ is weak,  in agreement with the pentamerization  argument. 

On the other hand, in the $n=2$ NA case, $C_{23}$ (spin without field and spin with field $h_2$, top left) is clearly significant 
in the $|M|=N/4$ plateaus emerging for small $|h_2|$, 
and small or zero in the same plateaus emerging for small $|h_1|$ and strong $|h_2|$, supporting the trimerization argument. 
This is  verified in $C_{24}$ (bottom left),  which is also  significant (zero) when $C_{23}$ is large (small) 
in these plateaus, entailing essentially no entanglement between trimers.  $C_{24}$ is also 
non-negligible at the $M=0$ plateau, where nearest spins with no field become entangled  due to the field induced frustration. 
It is also confirmed that all concurrences are finite at the $|M|=Ns-1$ band, 
in agreement with Eq.\ (\ref{Cij}).

\section{Conclusions \label{III}}

We have shown that 
$n$-alternating field configurations can lead to novel GS phase diagrams which differ  significantly  from those of 
the standard  alternating case. They can exhibit  non-trivial magnetization plateaus associated  
with field induced frustration and polymerization phenomena, which  persist for large sizes   
as verified by DMRG calculations. These plateaus 
satisfy a quantization rule compatible with the OYA criterion and are shown to stem from field induced polymers 
with definite magnetization, where spins trapped between spins with fields become highly entangled among themselves 
but are essentially disentangled with spins in another polymer.
Exact analytic expressions for the boundary  in field space of the fully aligned phase, valid for all $n$,  
were also derived, and imply a critical $n$-dependent 
anisotropy  $j_z^c(n)/j$ below which  the aligned phase together with a whole interval of GS magnetizations 
become unreachable even for arbitrarily strong fields. The boundary of the aligned phase represents 
in addition the onset of GS entanglement (as well that of the symmetry-breaking phase at the mean field level), 
with pairwise entanglement acquiring there full range. 
 These results open new possibilities for applications of finite chains with simple interactions under controllable fields, 
 such as entanglement tuning and plateaus formation at rational values of the scaled magnetization, and 
pave the way to study the emergence of critical phenomena induced through non-uniform fields within more general 
architectures and couplings.  

\acknowledgments
We thank Dr. J. M. Matera for useful discussions. The authors acknowledge support from CONICET (MC, NC, CAL) 
and CIC (RR) of Argentina. 	%Work supported by CONICET PIP 11220150100732.
\appendix
\section{Border of the aligned phase in the $n$-alternating spin-$s$ $XXZ$ system}
We first prove  Eqs.\  (\ref{alph})--(\ref{beta}). 
In the standard alternating case $n=1$, $\Delta H_n$ in (\ref{Det})  is just a $2\times 2$ matrix, 
\begin{equation}\Delta H_1=\begin{pmatrix}
h_1+j_z&-j\\-j&h_2+j_z\
\end{pmatrix},\label{A1}\end{equation}
and a trivial calculation yields $a_1=1$, $b_1=j_z$ and $c_1=j_z^2-j^2$ in (\ref{dan}), with $\alpha_1=j$, $\beta_1=j_z$ 
(Eq.\  (\ref{cz})). In this case the lowest eigenvalue of $\Delta H_n$ is just  $\lambda_0(1)=j_z+\frac{h_1+h_2}{2}-\sqrt{(\frac{h_1-h_2}{2})^2+j^2}$, 
and Eq.\ (\ref{crit}) can be directly obtained from the condition $\lambda_0(1)>0$. 

For general $n\geq 2$, evaluation of ${\rm Det}[\Delta H_n]$ in Eq.\ (\ref{dan})  yields
\begin{equation}a_n=(d_{n-1})^2\,,\;\;b_n=d_{2n-1}\,,\label{an}
\end{equation} 
and  $c_n=d_{2n}-\frac{j^2}{4}d_{2n-2}-2\frac{j^{2n}}{4^n}$, 
where 
\begin{equation}
d_n=\begin{vmatrix}
j_z&-j/2&0&0&\ldots\\
-j/2&j_z&-j/2&0&\ldots\\
&&\ddots&&\\&&&\ddots&\\
0&\ldots&0&-j/2&j_z
\end{vmatrix}\,,\end{equation} is the determinant of an $n\times n$ 
Toeplitz \cite{BG.05} tridiagonal matrix $M_n$  
of elements 
$j_z\delta_{ij}-\frac{j}{2}\delta_{i,j\pm 1}$. 
It then satisfies  
\begin{equation}d_{n+1}=j_z d_{n}-(j/2)^2d_{n-1}\,,\end{equation}
for $n\geq 1$, with $d_1=j_z$, $d_0\equiv 1$, i.e., $(^{d_{n+1}}_{\;d_{n}})=A^n(^{j_z}_{1})$, 
with $A=(^{j_z\;-j^2/4}_{1\;\;\;\;\;\;\;0})$. Hence, for any $n\geq 1$, diagonalization of $A$, 
which has eigenvalues $\frac{1}{2}(j_z\pm h_s)=\frac{1}{2}j e^{\pm\gamma}$, with $h_s=\sqrt{j_z^2-j^2}=j\sinh\gamma$ and $\cosh\gamma=j_z/j$, leads to 
\begin{equation}
d_n=\frac{(j_z+h_s)^{n+1}-(j_z-h_s)^{n+1}}{2^{n+1} h_s}=\frac{j^n}{2^n}\frac{\sinh[(n+1)\gamma]}{\sinh\gamma}\,.\label{dn}\end{equation}
 Eqs.\ (\ref{an})--(\ref{dn}) then lead to 
$a_n=(\frac{j}{2})^{2n-2}\frac{\sinh^2 n\gamma}{\sinh^2\gamma}$,  $b_n=(\frac{j}{2})^{2n-1}\frac{\sinh2 n\gamma}{\sinh\gamma}$ and 
 $c_n=4(\frac{j}{2})^{2n}\sinh^2n\gamma$, implying 
$\alpha_n=2(\frac{j}{2})^{n}/d_{n-1}$, i.e.\ Eq.\ (\ref{alph}), with  $\beta_n=b_n/a_n$ given by (\ref{beta}). 
\qed 

Now, it is apparent from (\ref{dan}) and previous expressions  that the matrix $\Delta H_n$ is positive definite for $j_z>j$ and positive fields $h_1,h_2$ 
($a_n>0, b_n>0, c_n>0$ $\forall$ real $\gamma$). 
On the other hand, at the threshold value (\ref{critz}),  $\gamma=\imath\pi/n$ and Eqs.\ (\ref{an})--(\ref{dn}) 
lead to $a_n=b_n=c_n=0$, i.e.\ ${\rm Det}[\Delta H_n]=0$ $\forall$ $h_1,h_2$, indicating the presence 
of a {\it vanishing} eigenvalue of $\Delta H_n$ and hence the loss of stability of the aligned $M=Ns$ GS.  

The eigenvalues of $\Delta H_n$ represent of course  excitation energies constructed from single spin excitations when $\Delta H_n$ is positive  definite.  
The eigenvalue equation ${\rm Det}[\Delta H_n-\lambda \mathbb{1}]=0$ 
can   be  explicitly obtained  from  Eq.\ (\ref{dan}) and the previous expressions for $a_n$, $b_n$, $c_n$, 
replacing $j_z\rightarrow j_z-\lambda$ and $\gamma\rightarrow \imath\phi$: It reads 
\begin{equation}%\left(\frac{j}{2}\right)^{2n-2}
\frac{\sin^2 n\phi}{\sin^2 \phi}
[h_1h_2+\frac{j(h_1+h_2)\sin\phi}{\tan n\phi}-j^2\sin^2\phi]=0
\,,\label{eig}\end{equation}
where $\cos\phi=(j_z-\lambda)/j$. It is first seen that (\ref{eig}) is fulfilled  for $\phi=\pi k/n$, $k=1,\ldots,n-1$,
implying the $n-1$ {\it field-independent} eigenvalues  
\begin{equation}\lambda_k(n)=j_z-j\cos(\pi k/n), \;\;k=1,\ldots,n-1\,.\label{lk1}\end{equation} 
The lowest one, $\lambda_1(n)$, vanishes precisely at the threshold (\ref{critz}), becoming negative for $j_z<j_z^c(n)=j\cos(\pi/n)$. 
In addition, the bracket in (\ref{eig}) leads to the remaining  $n+1$ {\it field-dependent} eigenvalues.  
The lowest one is obtained for $\phi=\phi_0<\pi/n$ (and $\phi_0>0$), leading to
\begin{equation}\lambda_0(n)=j_z-j\cos\phi_0<\lambda_1(n)=j_z-j\cos\pi/n\,,\end{equation}
with equality approached only at strong fields $h_1, h_2\gg j$ 
(where $\phi_0\approx \pi/n-\frac{h_1+h_2}{nh_1h_2}\sin(\pi/n)$, approaching  $\pi/n$ for $h_1,h_2\rightarrow+\infty$). 
Thus, for $j_z>j_z^c(n)$, $\Delta H_n$ is always positive definite at sufficiently strong fields ($\lambda_0(n)>0$), 
while for $j_z\leq j_z^c(n)$, it is non-positive ($\lambda_0(n)<0$) at all finite fields and the aligned state can no longer be a GS. 

Replacing $j_z\rightarrow j_z-\lambda$ in (\ref{dn}), it is also seen that the eigenvalues of the $(n-1)\times (n-1)$ 
Toeplitz matriz $M_{n-1}$  are just those of Eq.\ (\ref{lk1}) \cite{BG.05}. 
This matrix is just the block of $(\Delta H)_n$ associated with the  $n-1$ contiguous spins with no field, 
which become decoupled from the aligned spins with field for $h_1,h_2\rightarrow \infty$.  Hence, 
$-j\cos\pi/n$ represents the lowest energy of the $n-1$ spins trapped between the two aligned spins at  $j_z=0$ and magnetization  $(n-1)s-1$, 

While a positive definite matrix $\Delta H_n$ is in principle  a necessary 
condition for stability of the $M=Ns$ GS, it turns out to be sufficient for $h_1+h_2>0$, 
since in this case the GS magnetization decreases in steps of length $1$ from its maximum $M=Ns$   
as the fields $h_1,h_2$ decrease from $+\infty$  (Fig.\ \ref{f3}).  The only exception occurs  
for $j_z>j$ along the line  $h_1+h_2=0$ between the factorizing fields (see bottom panels in Fig. 3), 
where the aligned states $M=\pm Ns$ become degenerate  GS's if $|h_i|<h_s$,  and all GS magnetizations plateaus merge if $|h_i|=h_s$.  

Finally, we note that in the mean field approximation, 
the onset of the symmetry-breaking phase is again determined by the fields where the matrix $\Delta H_n$ ceases to be positive definite, 
since it is constructed from single spin excitations. A symmetry-breaking product 
state $|\Psi_{\rm mf}\rangle\propto e^{-\imath \sum_i \theta_i S^y_i}|M=Ns\rangle$ becomes in fact 
$\approx |Ns\rangle+\sum_i w_i|W_i\rangle$ for small $\theta_i$, 
with $w_i=\theta_i\sqrt{sK/2}$. Hence, a non-positive $\langle \psi_{\rm mf}|\Delta H|\psi_{\rm mf}\rangle$ 
is then equivalent to $\Delta H_n$ not being positive definite. 

\section{Reduced states and entanglement in the $M=Ns-1$ GS} 
The $|M=Ns-1\rangle$ GS will have the form 
\begin{equation} |Ns-1\rangle=\sum_{i=1}^{2n} w_i|W_i\rangle\,,\label{Ns1}
\end{equation}
where $|W_i\rangle$ are the states (\ref{W}) and the coefficients $w_i$ are obtained from the 
diagonalization of the  matrix $\Delta H_n$ of elements (\ref{Hn}) ($\sum_i|w_i|^2=1$, with $w_i>0$ $\forall$ $i$ for $J>0$). 
 From  the form (\ref{W}) of the states $|W_i\rangle$,  it becomes apparent that the reduced state 
 $\rho_{kl}={\rm Tr}_{\overline{kl}}|Ns-\!1\rangle\langle Ns-\!1|$ of any two distinct spins 
$k\neq l$ in the state  (\ref{Ns1}) will depend just on their positions $i,j$ within the cell each spin belongs, 
but not on their absolute distance $|k-l|$. Since the reduced state  will also commute with the total spin $S^z_{kl}=S^z_k+S^z_l$ 
of the pair, it will be given, for $M=Ns-1$,  by ($K=N/2n$ is the number of cells) 
\begin{equation}\rho_{ij}=\begin{pmatrix}
1-\frac{|w_i|^2+|w_j|^2}{K}&0&0&0\\0&\frac{|w_j|^2}{K}&\frac{w_jw_i^*}{K}&0\\0&\frac{w_iw_j^*}{K}&\frac{|w_i|^2}{K}&0\\0&0&0&0\end{pmatrix}
\label{rij}\,,\end{equation}
in the subspace spanned by the states    $\{|ss\rangle,|s,s-1\rangle,|s-1,s\rangle,|s-1,s-1\rangle\}$, 
where $|s-1\rangle=\frac{1}{\sqrt{2s}}S^-|s\rangle$. Eq.\ (\ref{rij}) is valid for any $s$ and $i,j=1,\ldots,2n$. It can then be always  
considered as a mixed state of an {\it effective  two-qubit system}, as just states $|s\rangle$ and $|s-1\rangle$ 
 are involved at each spin. A similar expression holds for the reduced state in the  $M=-Ns+1$ GS in the corresponding subspace.  

The state (\ref{rij}) is a mixed state with two non-zero eigenvalues $p_{ij}=(|w_i|^2+|w_j|^2)/K$ and $1-p_{ij}$. 
Its entropy $S(\rho_{ij})=-{\rm Tr}\,\rho_{ij}\log_2\rho_{ij}$ is the entanglement entropy of the pair  
with the rest of the chain. On the other hand, the entanglement between both spins  
can be measured through its entanglement of formation \cite{BD.96},  defined as the convex roof extension of the pure state entanglement entropy: 
For a general mixed state $\rho\equiv \rho_{AB}$, it is the minimum of the average entanglement over all decompositions of $\rho$ as convex mixture of pure states: 
\begin{equation} 
E_f(\rho)=\mathop{\rm Min}_{\{q_\alpha,|\Psi_\alpha\rangle\}}\sum_\alpha q_\alpha E(|\Psi_\alpha\rangle)\,,
\label{Eof}
\end{equation}
where $\sum_\alpha q_\alpha|\Psi_\alpha\rangle\langle\Psi_\alpha|=\rho$, $q_\alpha\geq 0$, $\sum_\alpha q_\alpha=1$, and 
$E(|\Psi_\alpha\rangle)=S(\rho_A^{\alpha})=S(\rho_B^{\alpha})$ is the entanglement entropy of  $|\Psi_\alpha\rangle$ ($\rho_{A(B)}^{\alpha}={\rm Tr}_{B(A)}|\Psi_{\alpha}\rangle\langle\Psi_\alpha|$ are the reduced states). 

While the evaluation of Eq.\ (\ref{Eof})  in the general case is a computationally hard problem, 
for a two-qubit mixed state $\rho$ it can be analytically determined through the concurrence $C(\rho)$ \cite{WW.97}, 
defined as in Eq.\ (\ref{Eof}) with  $E(|\Psi_{\alpha}\rangle)\rightarrow
C(|\Psi_{\alpha}\rangle)=\sqrt{S_2(\rho_A^\alpha)}=\sqrt{S_2(\rho_B^\alpha)}$, where  
$S_2(\rho)=2(1-{\rm Tr}\,\rho^2)$ is the linear entropy. For a two-qubit state $\rho$ the concurrence can be calculated as  \cite{WW.97}
\begin{equation}
C(\rho)={\rm Max}[2\lambda_{\rm max}-{\rm  Tr}\,R,0]\,, \,\,\,\, R=[\rho^{1/2}\tilde{\rho}\rho^{1/2}]^{1/2},
\label{conc}
\end{equation}
where $\lambda_{\rm max}$ denotes the largest eigenvalue of $R$ 
and $\tilde{\rho}=\sigma_y\otimes\sigma_y
\rho^*\sigma_y\otimes \sigma_y$ is the spin flipped density, with $\sigma_y$ the Pauli matrix. Eq.\ (\ref{Eof}) then becomes \cite{WW.97} 
\begin{equation}
E_f(\rho)=-\sum_{\nu=\pm} q_{\nu} \log_2 q_{\nu}\,,\;\;q_{\pm}=\frac{1\pm \sqrt{1-C^2(\rho)}}{2}\,, 
\end{equation} 
and is just an increasing convex function of $C(\rho)$, with 
$E_f(\rho)=C(\rho)=1$ $(0)$ for a maximally entangled (separable) two-qubit state. For  a pure  state 
$\rho=|\Psi\rangle\langle\Psi|$,  $C(\rho)=\sqrt{S_2(\rho_{A(B)})}$  and $E_f(\rho)$  
becomes the standard entanglement entropy $S(\rho_{A(B)})$. The concurrence is itself a proper entanglement monotone 
\cite{GV.00}  and satisfies a  monogamy inequality \cite{CKW.00,OV.06}.     
 
In the case of the state  (\ref{rij}), the pair  concurrence $C_{ij}=C(\rho_{ij})$ obtained from Eq.\ (\ref{conc}) 
 becomes just $C_{ij}=2|(\rho_{ij})_{23}|=2|w_i w_j|/K$ and is then given by Eq.\ (\ref{Cij}). 
These concurrences saturate the monogamy inequality, 
 namely 
\begin{equation}\sum_{l\neq i} C_{il}^2=4\frac{|w_i|^2}{K}\left(1-\frac{|w_i|^2}{K}\right)=C_{i,\text{rest}}^2\,,\end{equation} 
where $C_{i,\text{rest}}^2=S_2(\rho_i)=2(1-{\rm Tr}\,\rho_{i}^2)$  is the tangle of single spin $i$ with the rest of the chain, with  
\begin{equation}\rho_i=\begin{pmatrix}1-|w_i|^2/K&0\\0&|w_i|^2/K\end{pmatrix}\,,
\end{equation}
the reduced state of spin $i$ in the state (\ref{rij}).  For a general state we have instead 
$\sum_{l\neq i} C_{il}^2\leq C_{i,{\rm rest}}^2$ \cite{CKW.00,OV.06}.  

While Eq.\ (\ref{Cij}) is  valid for any spin $s$ due to the form (\ref{rij}) of the reduced pair state, in general states 
the pairwise entanglement of formation for spin $s\geq 1$ will not be analytically computable. Instead, we can use as 
computable quantifier the negativity $N(\rho)$ \cite{VW.02}, defined as the absolute value of the sum of the negative 
eigenvalues of the partial transpose of $\rho\equiv \rho_{AB}$. According to the Peres criterion \cite{PC.96}, $N(\rho)>0$ 
implies entanglement (though the converse does not hold in general). In the $|M|=Ns-1$ region, the negativity $N_{ij}=N(\rho_{ij})$ 
determined  by the state  (\ref{rij}) is, 
 setting   $\gamma_{ij}=1-(|w_i|^2+|w_j|^2)/K$,
\begin{equation}
N_{ij}=\frac{1}{2}\left(\sqrt{\gamma_{ij}^2+4|w_i|^2|w_j|^2/K^2}-\gamma_{ij}\right)\,,
\end{equation}
with $N_{ij}\rightarrow C^2_{ij}/2$ for large $K$. 

Due to the symmetry $w_{n+1+i}=w_{n+1-i}$ valid for $i=1,\ldots,n-1$ in the exact GS under cyclic conditions, 
the coefficients $w_i$ in (\ref{Ns1}) can be obtained by diagonalizing an effective $(n+1)\times (n+1)$ matrix $\Delta H'_n$. 
Altogether there are just $(n+1)$ distinct coefficients $w_i$ and hence just $(n+1)(n+2)/2$ distinct pairwise concurrences 
and negativities for general $h_1,h_2$ in the $|M|=Ns-1$ GS. 

\section{Exact solution of the $XX$ chain in $n$-alternating field configurations}
When $J_z=0$,  the $XXZ$ model  reduces to the $XX$ model. For $s=1/2$, the ensuing  Hamiltonian can be mapped exactly 
to a bilinear fermionic form in the annihilation $c^\dagger_j$ and creation $c_j$ operators by means of the Jordan-Wigner transformation \cite{LSM.61} $
c^\dagger_j =S^+_j \exp(-\imath\pi\sum_{k=1}^{j-1}S^+_k S^-_k)$ for each value of the fermionic number parity (i.e., the $S_z$-parity)
\begin{equation}
P\equiv\exp(\imath \pi \mathbf{N})=\sigma=\pm1\,, \label{par}
\end{equation}
 where $\mathbf{N}=\sum_{j=1}^N c^\dagger_j c_j= S^z+N/2$ is the fermion number operator. This leads to 
\begin{equation}H=-\sum_j[h_j (c^\dagger_j c_j-1/2)-\eta_j^\sigma\frac{J}{2}(c^\dagger_{j+1}c_j+c^\dagger_jc_{j+1})]
\end{equation}
where, for cyclic conditions, $\eta_j^-=1$ $\forall\, j$ and $\eta_j^+=1$ ($-1$) for $j\leq N-1$ ($j=N$). After a discrete Fourier 
transform of the fermion operators, it can be expressed as a sum of $K$ $2n\times 2n$ matrices ${\bf H}_k$:
\begin{align}
H &=-\sum_{k=1-\delta_{\sigma 1}/2}^{K-\delta_{\sigma 1}/2}{\bf c}'^\dagger_k \cdot {\bf H}_k {\bf c}'_k-\epsilon\,,\;\; \label{Hf2}\\
{\bf H}_k&= 
\begin{pmatrix}
h^+ + J\cos\omega_k & h^- &\ldots \\
h^- & h^+ + J\cos(\omega_k+\frac{\pi}{n}) &\ldots  \\
\vdots& \vdots&\ddots
\end{pmatrix}\!\!, \label{Hk} \\
&= {\bf D}_{k} +{\bf A}\,, 
\label{Hk2}
\end{align}
with ${\bf c}'^\dagger_k=(c'^\dagger_k,c'^\dagger_{k+N/(2n)},\ldots,c'^\dagger_{k+(2n-1)N/(2n)})$, ${\bf D}_{k}$ a diagonal matrix of 
elements $({\bf D}_{k})_{ii}= J\cos(\omega_k+\frac{\pi(i-1)}{n})$, ${\bf A}$ a circulant matrix specified by the vector 
$(h^+,h^-,h^+,h^-,\ldots)$, and 
\begin{equation}
h^\pm=\frac{h_1 \pm h_2}{2n}\,, \;\; \epsilon\!=\!\frac{N h^+}{2}\,, \;\; \omega_k=2\pi k /N\,. \label{heps} 
\end{equation}
Eq.\ (\ref{Hf2}) shows that the Fourier transformed $n$-alternating field configuration leads to off diagonal hopping terms 
specifying the allowed momentum values. The index $k$ is half-integer (integer) for $\sigma=1$ ($-1$). 

Due to the parity dependence of the energy levels, the number of GS magnetization transitions is associated to the number of 
times the single particle energies change sign \cite{CR.07}. Hence, field values at which single particle energies vanish 
can be determined by solving 
\begin{equation}
{\rm Det}\,[{\bf H}_k]=0\,,
\label{detz}
\end{equation}
with $k=1/2,1,\ldots,K$. 

For standard alternating fields $n=1$, Eq.\ (\ref{Hk}) becomes
\begin{equation}
{\bf H}_k= \left(
\begin{array}{cc}
h^+ +J \cos\omega_k  & h^-  \\
h^- & h^+ -J \cos\omega_k    \\
\end{array}
\right)\,,
\label{Hkalt}
\end{equation}
yielding the well known single particle energies \cite{JP.75,KO.90,CR.07,DE.08,DD.08,PA.08,DU.15}
\begin{equation}
_{1}\lambda_{k}^{\pm}=h^{+}\pm\sqrt{(h^-)^2+J^2\cos^2\omega_k}\,.    \label{spalt}
\end{equation}
In this case 
\begin{equation}
{\rm Det}\,[{\bf H}_k]= h_1h_2-J^2\cos^2\omega_k\,,
\label{DetAlt}
\end{equation}
and Eq.\ (\ref{detz}) determines $N/2$ hyperbolas in the $(h_1,h_2)$ field space, meaning that the GS will then exhibit 
definite magnetization plateaus ranging from $|M|=0$ to $|M|=N/2$. In particular, for $k=N/2$ the lowest $\sigma=-1$ 
parity level becomes negative and  we recover {\it exactly} the hyperbola $h_1 h_2=j^2$  of the $N/2\rightarrow N/2-1$ 
transition, in agreement with Eqs.\ (\ref{crit})--(\ref{crit2}) for $n=1$ and $j_z=0$.  For $n\geq 2$ 
the expressions for the eigenvalues are more involved.  

In the NA $n=2$ case,  the determinant of ${\bf H}_k$ is
\begin{equation}
{\rm Det}\,[{\bf H}_k]=\frac{J^4}{4}\sin^2 (2\omega_k)\,,\label{C11}
\end{equation}
which becomes zero only for $k=N/4$ and implies at least one identically zero single particle energy. 
The latter means that there is no single particle energy which changes sign as the fields are varied and 
indicates that there should be no GS magnetization transition. Furthermore, we now prove the following lemma:\\
{\bf Lemma 1}. {\it The GS of a finite  $XX$ spin system in a $n=2$ next-alternating field configuration is a 
	nondegenerate half-filled state with definite magnetization  $M=0$, $\forall\, h_1,h_2$.}

{\it Proof:} We first start by comparing the number of energy levels with negative single particle energies 
within each parity $\sigma$ and their ensuing lowest energy $E_\sigma$. Since ${\rm Det}\,[{\bf H}_k]
\geq 0$ $\forall k$ [Eq.\ (\ref{C11})]  
	then each matrix ${\bf H}_k$ is either positive (or negative) semi-definite, or it has two positive 
and two negative eigenvalues. However, since the 
determinant of any leading principal minor connecting $i$ with $i+2$ is $-J^2\cos^2(\omega_k)$, ${\bf H}_k$ 
cannot be positive nor negative semi-definite. 
In the $\sigma=1$ subspace, ${\rm Det}\,[{\bf H}_k]>0$  $\forall k=\{1/2,\ldots K-1/2\}$, entailing that 
there are always  $N/2=2K$ negative single particle 
energies, whereas for $\sigma=-1$ there are $N/2-1$, as one of the eigenvalues of ${\bf H}_{N/4}$ is 
identically zero. Due to this small, albeit important, 
difference in the number of negative energy levels, $E_1<E_{-1}$ $\forall\, h_1,h_2$. While this result 
can be numerically verified, for $h_2\!=\!\pm h_1\!=\!\pm h$ 
a series expansion of the energy difference between the lowest energies of  each parity, $\Delta E=E_{-1}-E_1$, 
shows that $\Delta E>0$   $\forall h$. 
Likewise, for strong fields a second order perturbation treatment in the couplings shows that the $M=0$ eigenstate 
is the GS $\forall J$. \qed


\begin{thebibliography}{999}
\bibitem{expl1}  The sign of $J$ can be changed by local rotations of angle $\pi$ around the $z$ axis at even (or odd) sites $j$.
\bibitem{expl2} These cases  are linked by  a  global  rotation of angle $\pi$ 
	around the $x$ or $y$ axis, which leaves the coupling unchanged.
\bibitem{DMRG} We usually perform 20 sweeps keeping up to 900 states.
\bibitem{expl3} In a pure state with definite total magnetization $M$ along $z$, the single spin reduced state $\rho_i$ 
	commutes with $S^z_i$ $\forall\,s$. For $s=1/2$ its eigenvalues are then $1/2\pm \langle S_i^z\rangle$,
	and its purity ${\rm Tr}\,\rho_i^2=\frac{1}{2}+2\langle S_i^z\rangle^2$.
\end{thebibliography}
\end{document}